\documentclass[]{spie}  
 
\usepackage{amsmath,amsfonts,amssymb}
\usepackage{graphicx}
\usepackage[colorlinks=true, allcolors=blue]{hyperref}
\usepackage{tikz,xcolor,hyperref}
\usepackage{wrapfig}
\definecolor{lime}{HTML}{A6CE39}
\DeclareRobustCommand{\orcidicon}{%
    \begin{tikzpicture}
    \draw[lime, fill=lime] (0,0) 
    circle [radius=0.16] 
    node[white] {{\fontfamily{qag}\selectfont \tiny ID}};
    \draw[white, fill=white] (-0.0625,0.095) 
    circle [radius=0.007];
    \end{tikzpicture}
    \hspace{-2mm}
}
\newcommand{\orcid}[1]{\href{https://orcid.org/#1}{\orcidicon}}

\title{Modeling of the diffuse background produced by the Vera C. Rubin Observatory M2 baffle scattered light}

\authorinfo{Further author information: E-mail: alessio.taranto@inaf.it}

\begin{document} 
\author[a,b]{Alessio Taranto\orcid{0009-0009-3271-3498}}
\author[a]{Gabriele Rodeghiero\orcid{0000-0002-3469-9863}}
\author[a,b]{Luca Rosignoli\orcid{0000-0002-0327-5929}}
\author[c]{Aashay Pai\orcid{0009-0008-9641-6065}}
\author[c,d,j,q]{Alex Drlica-Wagner\orcid{0000-0001-8251-933X}}
\author[g]{Elana K. Urbach\orcid{0000-0002-3205-2484}}
\author[e]{Fritz M\"uller}
\author[k]{Eli S. Rykoff}
\author[e]{Hannah M.M. Pollek\orcid{0009-0001-3368-4539}}
\author[f]{John Andrew}
\author[f]{Douglas R. Neill}
\author[f]{Parker Fragelius}
\author[h]{Tomislav Vucina}
\author[g]{Christopher W. Stubbs\orcid{0000-0003-0347-1724}}
\author[i]{Robert~H.~Lupton\orcid{0000-0003-1666-0962}}
\author[i]{Lee~S.~Kelvin\orcid{0000-0001-9395-4759}}
\author[h]{Kate Napier\orcid{0000-0003-4470-1696}}
\author[h]{Jacqueline Seron Navarrete}

\author[e]{Travis Lange\orcid{0009-0008-0596-4489}}
\author[k]{Andrew P. Rasmussen}
\author[k]{Aaron Roodman\orcid{0000-0001-5326-3486}}
\author[f]{Chuck F. Claver}
\author[k]{Joshua~E.~Meyers\orcid{0000-0002-2308-4230}}
\author[h]{Anastasia~Alexov\orcid{0009-0000-7835-3963}}
\author[i]{Keith Bechtol}
\author[h]{Brian~Stalder\orcid{0000-0003-0973-4900}}
\author[l]{R. Lynne Jones}
\author[h]{Leanne P. Guy \orcid{0000-0003-0800-8755}}
\author[h]{Tiago Ribeiro}
\author[h]{Erik Dennihy\orcid{0000-0003-2852-268X}}
\author[h]{Bruno C. Quint\orcid{0000-0002-1557-3560}}
\author[m]{Aaron~E.~Watkins\orcid{0000-0003-4859-3290}}
\author[h]{Alysha B. Shugart\orcid{0009-0000-6778-7168}}
\author[h]{Lukas Eisert}
\author[h]{Kevin Fanning}
\author[h]{Marina S. Pavlovich\orcid{0000-0001-5560-7051}}
\author[e,h]{Yijung Kang\orcid{0000-0002-5261-5803}}
\author[r,h]{Hye Yun Park\orcid{0000-0002-7295-2743}}
\author[h]{Paulo Lago}
\author[h]{Kris Mortensen}
\author[h]{Paulina Venegas Salas}
\author[h]{Minhee Hyun\orcid{0000-0003-4738-4251}}
\author[h]{Karla Peña Ramírez\orcid{0000-0002-5855-401X}}
\author[h]{David Sanmartim\orcid{0000-0002-9238-9521}}
\author[h]{Qianjun Hang}
\author[h]{Gonzalo Aravena\orcid{0009-0006-5850-4860}}
\author[h]{Kshitija Kelkar\orcid{0000-0002-8130-3593}}
\author[h]{Carlos A. L. Morales Marín\orcid{0000-0003-0203-3407}}
\author[h]{Danica Žilková\orcid{0000-0002-5726-3640}}
\author[h]{Eric J. Christensen\orcid{0009-0001-9424-2291}}
\author[i]{Yusra Alsayyad}
\author[h]{William O'Mullane\orcid{0000-0003-4141-6195}}
\author[n]{Enrico Giro\orcid{0000-0001-7301-8285}}
\author[o]{Rodolfo Canestrari\orcid{0000-0003-4591-7763}}
\author[h]{Sandrine J. Thomas\orcid{0000-0002-9121-3436}}
\author[e,h]{Kevin A. Reil\orcid{0000-0002-2234-749X}}
\author[h]{Claudio H. Araya Cortes}
\author[h]{Roberto Tighe}
\author[h]{Holger Drass\orcid{0000-0002-7790-9971}}
\author[h]{Pablo Zorzi}
\author[p]{Massimo Brescia}

\affil[a]{\small INAF OAS, Via Gobetti 93/3, I-40129, Bologna, Italy}
\affil[b]{\small Department of Physics and Astronomy, University of Bologna, Via Gobetti 93/2, I-40129, Bologna, Italy}

\affil[c]{\small Department of Astronomy \& Astrophysics, University of Chicago, Chicago, IL 60637, USA}
\affil[d] {\small Fermi National Accelerator Laboratory, Batavia, IL 60510, USA}

\affil[e]{\small SLAC National Accelerator Laboratory, Menlo Park, CA 94025, USA}
\affil[f]{\small Vera C. Rubin Observatory Project Oﬀice, 950 N. Cherry Ave., Tucson, AZ 85719, USA}
\affil[g]{\small Department of Physics, Harvard University, 17 Oxford St., Cambridge MA 02138, USA}
\affil[h]{\small NSF-DOE Vera C.\ Rubin Observatory / NSF NOIRLab, Casilla 603, La Serena, Chile}
\affil[i]{\small Department of Astrophysical Sciences, Princeton University, Princeton, NJ 08544, USA}
\affil[j]{\small Kavli Institute of Cosmological Physics, University of Chicago, Chicago, IL 60637, USA}
\affil[k]{\small Kavli Institute for Particle Astrophysics and Cosmology, SLAC National Accelerator Laboratory, Menlo Park, CA 94025,USA}
\affil[l]{\small University of Washington, Dept. of Astronomy, Box 351580, Seattle, WA 98195, USA}
\affil[m]{\small Centre for Astrophysics Research, University of Hertfordshire, College Lane, Hatfield AL10 9AB, UK}
\affil[n]{\small INAF OATS, Via Giovan Battista Tiepolo 11, 34143, Trieste, Italy}
\affil[o]{\small INAF IASF, Via Ugo la Malfa 153, 90146, Palermo, Italy}
\affil[p]{\small Department of Physics ”E. Pancini”, University Federico II of Napoli, Via Cintia, 80126 Napoli, Italy}
\affil[q]{\small NSF-Simons AI Institute for the Sky (SkAI),172 E. Chestnut St., Chicago, IL 60611, USA}
\affil[r]{Duke University, Science Dr, Durham, NC 27710, United States.}

\authorinfo{Further author information: E-mail: alessio.taranto@inaf.it}

\maketitle

\begin{abstract}
 
    The Vera C. Rubin Observatory, with its unprecedented field of view and fast focal ratio
    (f/1.2), will survey the entire sky every 3.5 nights. This unique capacity also requires dealing with off-axis light that can potentially produce stray light artefacts on the images. The secondary mirror (M2) cell baffle restricts the light that reaches the LSSTCam detector and it contributes to shaping the inner edge of the telescope optical pupil. This work studies the contribution to the background of the residual light scattered by the M2 baffle when illuminated by highly off-axis light beams. The evanescence of this feature, together with the challenge of isolating it from the sky background, led to the necessity of performing multi-wavelength in-dome tests using a Collimated Beam Projector (CBP), normally used for calibration purposes. To complete the analysis, in addition to the in-dome tests, an on-sky observational campaign was conducted. This campaign employed both stellar targets and the Moon as illumination sources in order to determine the actual energy associated with the feature.
    The test data have been retro-fitted thanks to the combination of ray tracing simulation and CBP/on-sky data to infer the intensity and spatial distribution of the background scattered light within the different LSSTCam filters. We quantified the on-sky impact of scattered light from the M2 baffle, both for light coming from bright and red stars and from the Moon. We also developed an approximate relation to transform the in-dome measurements into predictions of on-sky behavior. This transformation was achieved by comparing the illumination footprint produced by an off-axis stellar source with that generated by the Collimated Beam Projector (CBP), and by mapping the stellar Spectral Energy Distribution (SED) onto the CBP’s set of discrete monochromatic wavelengths. Finally, we extrapolated the scattered-light behavior characterized for the Moon to stellar sources in order to obtain a comprehensive description of the scattered-light contribution over the full range of source magnitudes.
    
\end{abstract}

    \section{Introduction}
Located in the Chilean Andes on the Cerro Pachòn mountain, near La Serena, the Vera C. Rubin Observatory (Rubin) is the latest big astronomical facility to come online, which acquired its first photon in October 2024 using the commissioning camera (ComCam) and then the first light with the scientific camera (LSSTCam) in April 2025. This single instrument telescope has a unique purpose: systematically exploring the night sky to produce the Legacy Survey of Space and Time survey (LSST), a 10 year survey able to capture not only a very wide portion of the southern sky, $30.000\mathrm{deg^2}$ but also its variations in time within six different photometric bands \textit{ugrizy} spanning from  $320\,\mathrm{nm}$ to  $1100\,\mathrm{nm}$ \cite{Ivezi__2019}. 

For survey telescopes, being able to survey big portions of the sky in a short time is crucial. This ability is driven by the etendue of the optical system; it is computed as the area of the entrance pupil times the solid angle a source subtends as seen from the pupil. The Rubin etendue is one order of magnitude higher than any other existing facility \cite{Ivezi__2019}. The etendue and the fast slewing speed of the telescope are achieved thanks to its innovative design, a three-mirror modified Paul-Baker, Mersenne-Schmidt system with a very fast $f/1.234$ \cite{marsenne} \cite{angeli}. The innovative design led to the construction of a bi-concave mirror with both the primary and the tertiary mirrors part of the same glass blank. The total diameter of the glass blank is 8.4m, where M1 corresponds to the concave outer annulus with an inner diameter of 5m, resulting in an effective radius of 6.7m, while M3 is concave as well and has a radius of 5m. This bi-concave mirror is often referred to as M1M3 \cite{sebag16}. The secondary mirror is a convex 3.45m diameter mirror, whose size is comparable to the next generation of secondary mirrors for the upcoming extremely large telescopes (ELT, TMT, GMT) \cite{m2gabri}.
Besides the mirrors, the other fundamental component of the telescope is the Legacy Survey of Space and Time Camera (LSSTCam), which is composed of three refractive elements, with the first lens (L1) having a diameter of $1.6\mathrm{m}$. The camera is, at the moment, the biggest digital camera ever built. It is equipped with 189  $4\mathrm{k}\times4\mathrm{k}$  science CCD detectors, resulting in an unprecedented resolution of 3200 Megapixels to sample a total on-sky area of $9.6 \mathrm{deg^2}$ in each observation. 

The Vera C. Rubin Observatory serves a very large portion of the astronomical community worldwide with the different science cases on which the 10-year-long survey will focus:
\begin{itemize}
    \item Probing dark energy and dark matter
    \item Taking an inventory of the Solar System
    \item Exploring the transient optical sky
    \item Mapping the Milky Way
\end{itemize}

Starting from the stringent scientific requirements to perform the observations, thousands of technical requirements for the system have been derived; these include operational (i.e. the effective derived image quality or revisit times), design (i.e. filter complement), safety, performance and many more. At the time of writing, the system is being optimized in preparation for the start of the LSST survey, which requires stable, constant, and reliable system operability. In practice, this requires delivering an image quality close to the atmospheric seeing limit while maintaining high astrometric and photometric precision.

A major challenge in large-aperture, wide-field facilities is the control and mitigation of stray light. Stray light is defined as any radiation that reaches the focal plane through non-nominal light paths \cite{10.1117/12.3113387}. Such radiation originates mostly from off-axis sources through direct or scattered paths \cite{gabri2026}. In direct paths, the light only travels on optical elements through a non-desired order, while scattered paths originate when light is scattered from non-optical surfaces and contaminates the optical beam. In this paper, we will focus on the latter. If not properly controlled, stray light increases detector noise, creates spurious features, reduces image contrast, and degrades the overall scientific performance. While the LSST incorporates detailed sky background models to account for scattered light from the Moon and twilight, mechanical suppression is typically achieved using baffles and covers that prevent direct illumination of critical surfaces for specified off-axis exclusion angles. Additional suppression is provided by internal vanes, which cast shadows to minimize the area simultaneously illuminated by external sources and visible from the detector \cite{10.1117/12.3063381}. Ideally, vane geometries employ sharp edge profiles and optimized spacing so that only the vane edges contribute to scattering paths \cite{mon_mirr}. Complementary mitigation is obtained through blackening treatments applied to non-optical surfaces in order to maximize absorptivity and minimize the Bidirectional Reflectance Distribution Function (BRDF). Although conventional black anodization is commonly used, its optical performance strongly depends on wavelength and surface finish, often becoming increasingly reflective in the near-infrared \cite{Marshall_2014}. Optionally, some advanced coatings can be considered in high stray light requirement demanding scenarios; for example, the Aeroglaze Z306 coating is often used in ground-based and aerospace applications as an industry standard \cite{mon_mirr}. In the case of the M2 baffle at the Vera C. Rubin Observatory, the surface treatment is a black anodization; its measured profile is shown in Figure \ref{fig:bafref}. A principle mitigator of stray light for the Rubin Observatory is the Light-Wind Screen (LWS). The LWS covers most of the dome's slit, except the actual area of the light path. This item was not yet installed during this investigation. 

Within the Rubin Observatory optical design, the M2 baffle structure is particularly critical, as it constitutes one of the main regions where illuminated surfaces may overlap with surfaces directly visible to the focal plane. The observatory incorporates detailed sky-background models to account for diffuse contributions from twilight and moonlight; however, mechanical stray-light suppression remains essential to prevent direct and multiply scattered illumination from off-axis sources. The final goal of installing a light baffle surrounding the secondary mirror is to shield the focal plane from unwanted light coming from inside the dome or from mild off-axis sources. 

This work aims to characterize the diffuse background \emph{coming from} the M2 baffle when off-axis light sources are present in the surroundings of the observed sky.  Quantifying this contribution is essential for accurately modeling the expected local background during nominal operations and for assessing its impact on targeted LSST science cases requiring precise background estimation and high photometric accuracy, such as Low Surface Brightness (LSB) science. During the Observatory design phase, an initial stray light study \cite{fred_06} was performed using different software and some complementary approaches with respect to the one carried out during the commissioning and this one \cite{gabri2026}. Already from the first technical light of Rubin, some stray light was detected \cite{2026arXiv260323786V}.
LSB science cases are likely to be among the most severely impacted by stray light contamination, as the studies of phenomena such as LSB-type dwarfs \cite{vanDokkum2015}, stellar haloes \cite{Wang2019, LSB} and intracluster light \cite{Montes2019} are almost always performed in the low count regime. Any small perturbation in the measured flux can lead to substantial systematic errors in the final results. In addition, background estimation is of crucial importance to avoid over- or under-subtracting any low surface brightness signal \cite{Kelvin2023}. Computed background models are not always aware of instrumental-driven contamination, and it is not always easy to map such contamination on sky. For these reasons, it is always better to mitigate the spurious contributions coming from stray light as much, and as early in processing, as possible.

The next chapters are organized as follows: section 2. \textit{Secondary mirror cell and baffle} describes the hardware structure of the M2 cell and its baffle whose stray light analysis will be the focus of section 3. \textit{Ray tracing simulations of M2 baffle scattered light}; in this section we will explain how the scattering process works and how the simulations suggest an observable effect, comparing also with the on-sky and in-dome tests. Section 4. \textit{In-Dome tests} describes the setups and the procedures of the set of tests carried out in an in-dome environment with the CBP, while section 5. \textit{On-Sky tests} describes the procedures adopted for the on-sky tests with different off-axis sources in different photometric bands. Section 6. \textit{Retrofitting and extrapolation} discusses the analytical procedures used to compare in-dome and on-sky test results, and it investigates the connection between results from the two scenarios and results from on-sky tests with different celestial bodies as light sources. Section 7. \textit{Results} shows the results obtained from the methods described in the latter section. Section 8 \textit{conclusions} discusses what the result implies in terms of image degradation and possible mitigation actions if required.

    \section{Secondary Mirror cell and light baffle} \label{SEC2}
The M2 system of the Vera C. Rubin Observatory is an important part of the Simonyi Survey Telescope. Due to the telescope’s very fast focal ratio of $f/1.234$, its wide $3.5$°  field of view, and its strict need for quick movement, the M2 assembly is essential for keeping precise optical alignment and reducing wavefront errors during observations.

The Rubin M2 mirror is a $3.4\mathrm{m}$ diameter, $100\mathrm{mm}$ thick Ultra Low Expansion (ULE) glass meniscus. It is one of the most advanced ground-based secondary mirrors, and it serves as a pathfinder for future extremely large telescopes (ELTs). The mirror is mounted in a welded steel cell with 78 electromechanical actuators, including 72 axial and 6 tangential actuators. These actively adjust support forces as the telescope moves, limiting the deformation of the glass and helping to keep the mirror’s optical surface stable under changing gravity and temperature conditions \cite{m2}.

\begin{wrapfigure}{r}{0.5\textwidth}
    \centering
    \vspace{-2pt}
    \includegraphics[width=0.5\textwidth]{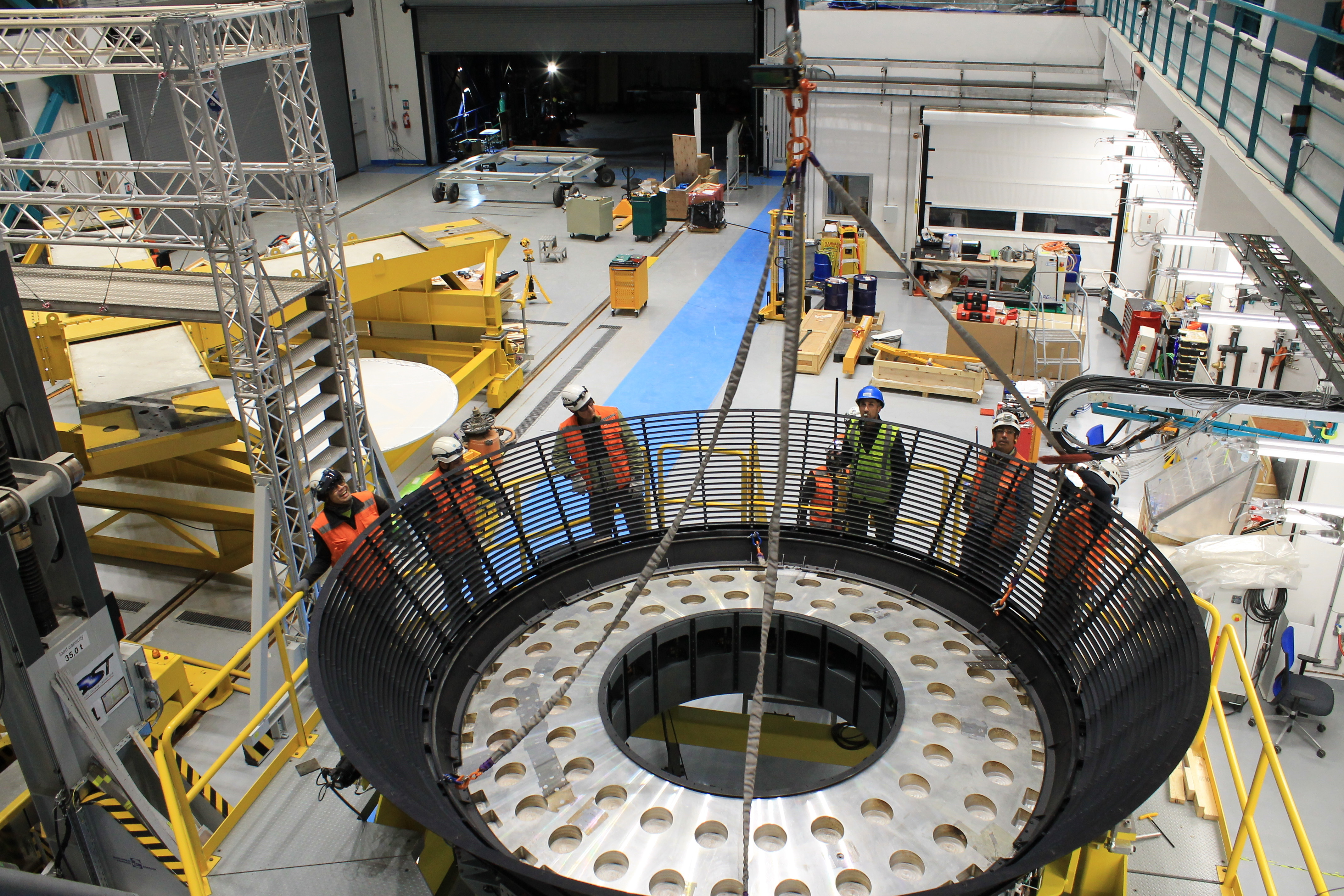}
    \caption{M2 baffle mechanical element mounted on the dummy mass representing M2 before being installed in its final location in the main telescope structure.}
    \label{fig:fotobaf}
\end{wrapfigure}

The M2 control system works in both open-loop and closed-loop modes. In closed-loop mode, the Force Balance system continuously adjusts internal forces to maintain a low-stress state for the mirror. It uses look-up tables for gravity and thermal effects, along with active optics bending-mode corrections from the AOS. Three of the axial actuators and three of the tangent links are normally disabled and function as hard points. These hard-points establish the mirror’s position relative to the cell, while load cells, encoders, temperature probes, and metrology systems continuously monitor its status and support conditions. This combined control and measurement system provides fast settling times and the optical stability needed for successful survey operations.

\begin{figure}
    \centering
    \includegraphics[width=0.6\linewidth]{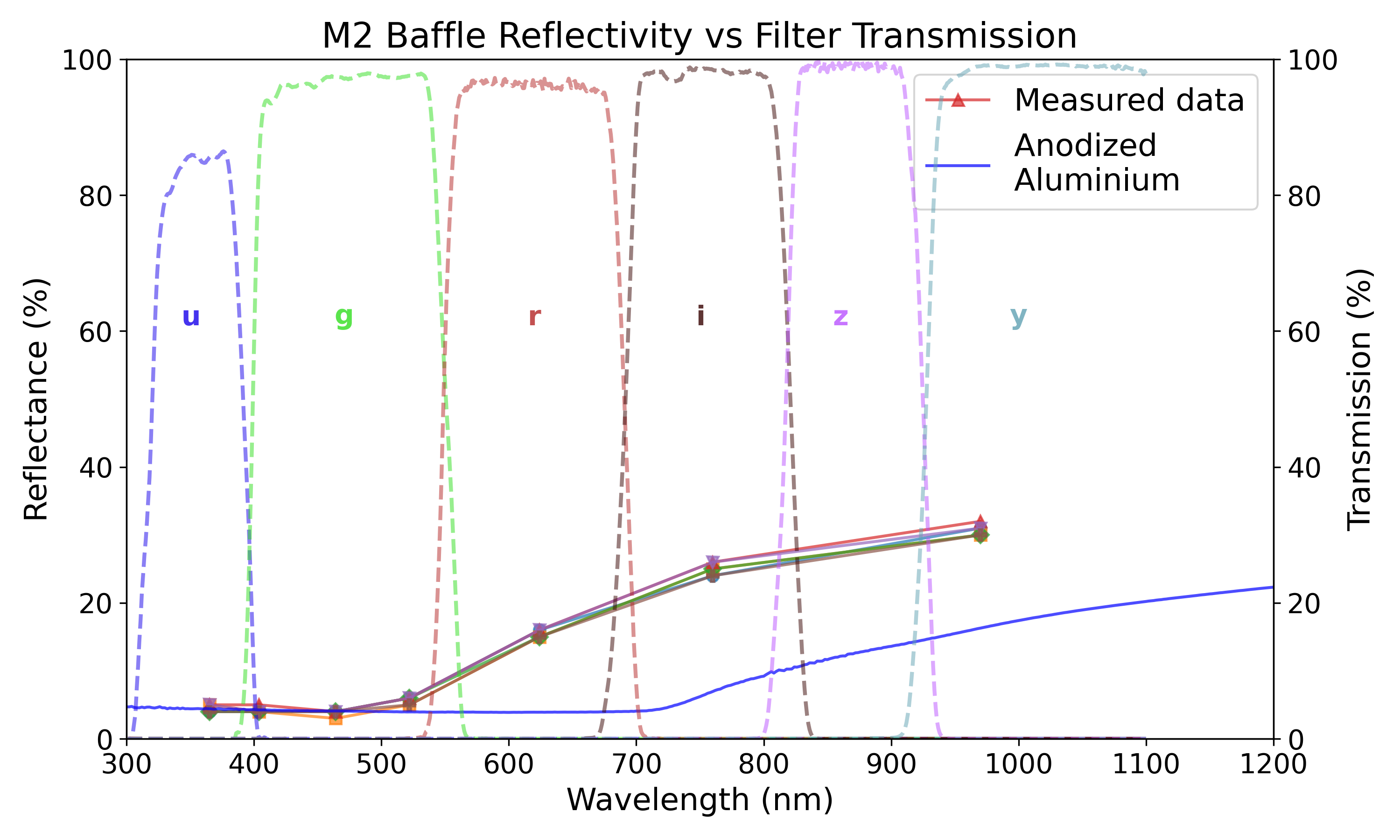}
    \caption{The solid curve connects the measured reflectances data points of the M2 baffle measured on-site in the different LSSTCam bands. The behavior is comparable to the expected inorganic treatment reference. }
    \label{fig:bafref}
\end{figure}

A critical element of the M2 assembly is the M2 baffle, which surrounds the secondary mirror assembly. It is a large, lightweight structure mechanically interfaced to the M2 cell and the telescope top-end assembly (TEA), composed of parallel circular vanes perpendicular to the optical path. This configuration suppresses stray light while preserving free air flow over the mirror surface, which is essential for minimizing mirror seeing. Its primary function is the suppression of stray light and multiple scattering paths that would otherwise reach the focal plane. In the Simonyi telescope optical configuration, large off-axis beams traverse the telescope structure and contributions from structural scattering can degrade photometric uniformity and background stability of acquired images. In this context, the M2 baffle acts to intercept light that would otherwise directly illuminate M3 and subsequently the detector. The vane-based geometry, introduced to maintain ventilation, creates potential double-bounce scattering paths between surfaces; this effect was mitigated through the use of black anodized aluminum. 
Although the theoretical single bounce reflectivity of this surface (~20\%) is relatively high for optical applications, the expected double-bounce contribution (~4\%), combined with the small fraction of rays capable of such interactions (~10\%), leads to an estimated total transmission of ~0.4\% through the structure, which was considered acceptable at the design stage. The M2 baffle is manufactured in black anodized aluminum, specifically a sulfuric \textit{black anodize per MIL-A-8625 Type II Class 2}, with a coating thickness of approximately $5\mathrm{\mu m}$ that provides corrosion resistance and a blackened appearance. However, on site measurements show that the assembled component exhibits non negligible and wavelength dependent reflectivity (Figure \ref{fig:bafref}), reaching values close to 40\% in the $y$ band. Furthermore, the scattering power of the M2 baffle surface finishing is quite low, contributing to an increase in infrared radiation reflection.

\section{Ray tracing simulation of M2 baffle scattered light}\label{SEC3}
The software tool used for the ray tracing is Zemax–Ansys Optics® \cite{zemax}. In particular, we exploited its non-sequential mode where it is possible to import mechanical CAD objects. This allows modeling the behavior of non-optical components when they are illuminated by the nominal beam or by off-axis sources as in the study case for this work. The approach followed during the early stages of this work was to gather the CAD models of many different mechanical components of the telescope main structure and include them in the baseline optical design. Thus we produced a simplified digital twin of the telescope itself in a Zemax–Ansys Optics® non-sequential framework. In particular, we included in the model the TEA mechanical structure, the M2 cell, M2 baffle and LSSTCam components (Figure \ref{fig:fotobaf}). 

Thanks to reflectivity measurements, shown in Figure \ref{fig:bafref}, we were able to associate some optical properties to the M2 baffle. Inside the simulation framework we treated the M2 baffle as if it were a mirror with a 30\% reflectivity and a combined Lambertian-specular scattering model. The latter is a standard model for scattering due to random isotropic surface roughness.

\begin{figure}
    \centering
    \includegraphics[width=0.315\linewidth]{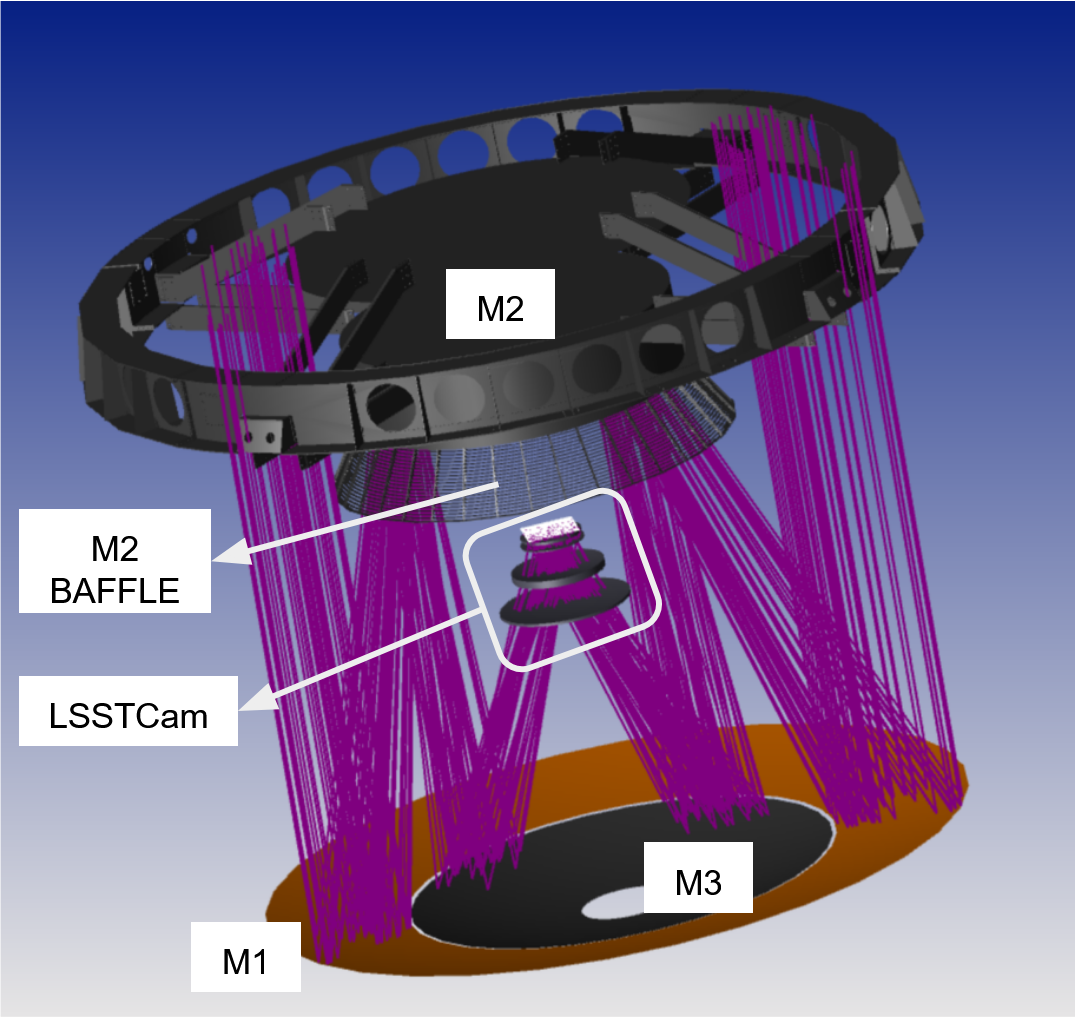}
    \hfill
    \includegraphics[width=0.3\linewidth]{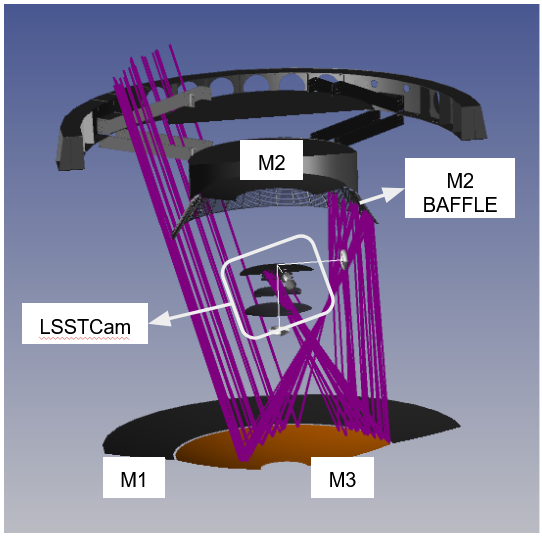}
    \hfill
    \includegraphics[width=0.34\linewidth]{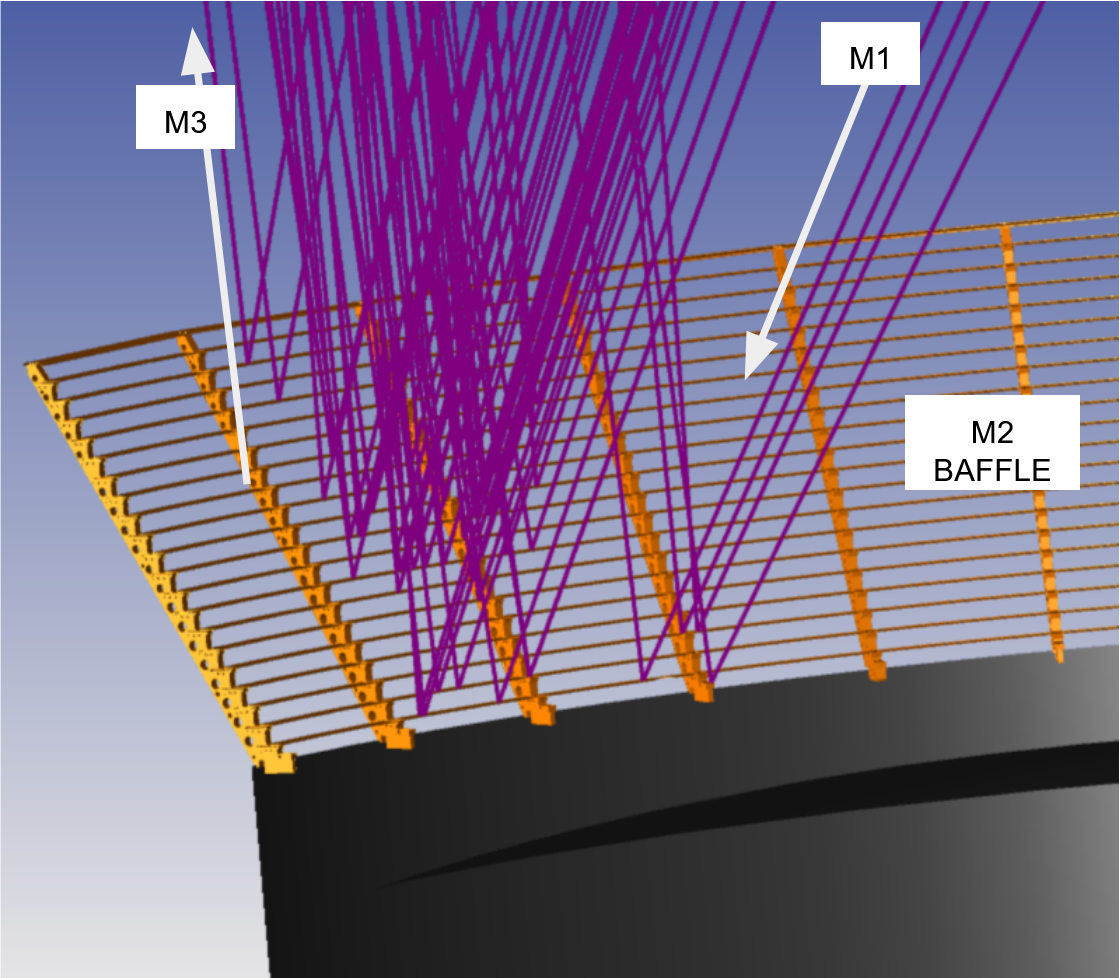}
    \caption{\textbf{Left:} The Zemax–Ansys Optics® digital twin of the Simonyi Survey Telescope with the mechanical components added upon the optical surfaces. The purple lines shows the primary optical path followed by the off-axis light that is responsible for the scattered light off the M2 baffle. The light footprint on M1 is shown in Figure \ref{fig:impact}. \textbf{Center:} The purple lines show the secondary optical path that skips M1 and has two reflections over M3, the artifact resulting from the combination of the two paths on the LSSTCam focal plane is shown in Figure \ref{fig:focalplane} \textbf{Right:} detailed view of the light path at the location where light from M1 hit the M2 baffle and bounce towards M3}.
    \label{fig:model}
\end{figure}

After the initial visual inspection in the dome during night operations, we started using the model to identify the possible off-axis angles at which the scattered light from the M2 baffle was able to effectively end up in the focal plane of LSSTCam. This step consisted of designing a realistic off-axis light source and then positioning it at many different off-axis angles to verify how the light scattered off the M2 baffle behaves. This rastering procedure resulted in a probable range of off-axis angles where the feature should materialize.  We used this hint from the simulation to design the in-dome and on-sky test campaigns and later to verify the CBP findings.

A highly effective capability for identifying the origin of specific stray light artifacts is the use of the filter string tool in single-ray optical path analysis \cite{string_zem}. Filter strings allow users to extract and display rays that follow defined propagation and scattering routes; with this analytical function, one can select rays that either hit or bypass specified objects and that undergo scattering interactions or generate ghost images. This specific feature helped us isolate the proper light path shown in Figure \ref{fig:model}. Thanks to this ray path analysis a secondary light path able to illuminate the M2 baffle was identified.  In this case, the off-axis light is skipping the primary mirror and lands directly on M3, thus following the path M3-M2 baffle-M3-LSSTCam. Although less common, this secondary path holds a comparable intensity to the main stray light path. This is because a large portion of the rays impinging on M1 are then dispersed away and do not reach the focal plane, while in the case of this secondary path the surviving probability of those rays is higher. In simulations we can isolate single paths and see the figures they produce on the focal plane. while this is impossible to do on-sky and with the CBP so in real images we see just the combination of the different paths illuminating the M2 baffle.

\section{In-Dome tests}\label{SEC4}
The characterization of the telescope optical transmission function is carried out by different calibrations, both in-dome and on-sky. Aside from the Simonyi Survey Telescope, the Observatory has an auxiliary telescope (AuxTel) with a primary mirror of 1.2m, located $\approx300\mathrm{m}$ away from the main telescope. It is equipped to acquire both images and spectra \cite{auxtel}.
Regarding the in-dome calibration, the Observatory has an annular flat field screen with an inner radius of $4.3\mathrm{m}$ and an outer radius of $9.3\mathrm{m}$, and a Collimated Beam Projector (CBP); the latter is used to project calibrated light fluxes in the telescope pupil to characterize the optics and detector response. The optical telescope assembly used for the CBP is a wide-field $\approx$30 cm diameter telescope able to produce a $\approx$24cm collimated beam and to illuminate a portion of the telescope entrance pupil at any user-defined positions \cite{cbp}. In order to study the transmission function over all the photometric filters mounted on the LSSTCam, the CBP is equipped with a monochromatic laser source and a set of different pinhole masks projecting different image spot asterisms.

For the study described in this work, we used the capability of the CBP to point towards the main telescope with any off-axis angle in a very stable and reproducible way. This aspect, in addition to the reduced light background we experience inside the dome, sets the best conditions for the characterization of the scattered light coming from the M2 baffle. Having both the light source and the M2 baffle inside the dome itself lets us exclude a wide range of other potential image contaminants that would complicate isolating the M2 baffle’s effect,suchg as thin high-altitude clouds, galactic cirrus, dusty sky regions and stray light from sunset or dawn.

In-dome tests have been performed at three different angles off-axis, $12$°, $16$° and $20$°, to sample the full range where background contamination is expected, as suggested by the ray-tracing simulations. Each test starts by pointing the dome and the telescope assembly to the parking position. Then, some safety checks of dome light, laser functionality, and mirror clearance are performed. Once the system functionality has been checked, the telescope and the CBP are co-pointed, so that the CBP illuminates a portion of M1M3 with a collimated beam aligned with the telescope optical axis. This step is done in order to confirm the correct functioning and pointing of the CBP and to verify that the correct mask is selected. We perform all of our CBP tests using a single 1 mm pinhole mask.

\begin{table}[t]
    \centering
    \begin{tabular}{|c|cc|cc|cc|cc|}
        \hline
              & \multicolumn{2}{c|}{Co-point} & \multicolumn{2}{c|}{12} & \multicolumn{2}{c|}{16} & \multicolumn{2}{c|}{20} \\\
                  & Az    & El   & Az    & El & Az & El & Az & El \\
        \hline 
        Telescope & 269°  & 20°  & 269°   &  25°  &  269°  & 20°  & 269°   &  20°  \\
        CBP       & -20.5°& 0.1° &  12°  &  -28°  & 13°   &  -29°  & 12°  & -32.5°   \\
        \hline
    \end{tabular}
        \vspace{2mm}

    \caption{Azimuth and elevation coordinates commanded to the telescope and the CBP to produce the desired test configurations at different off-axis angles. There is a degeneracy between the movement of the telescope and the CBP; we keep the CBP fixed in the co-aligned position and move the telescope to produce the desired off-axis angle.}
    \label{tab:placeholder}
\end{table}

Since the laser source in the CBP is capable of producing nearly monochromatic variable light, we perform all the in-dome tests without any filter mounted in the LSSTCam. LSSTCam filters are designed to have zero optical power, so performing the CBP tests without any filter is not affecting the final images in terms of opticalpath properties. This also helps to increase the throughput of the system, since we reduce the total number of optical elements. To sample the full spectral coverage equivalent to different LSSTCam filters, during the tests we project monochromatic light several times at intervals of $10\mathrm{nm}$ from $720\mathrm{nm}$ up to $1020\mathrm{nm}$, covering the $i$, $z$ , and $y$ bands. In general, the effect we are looking for is expected to be quite faint; for that reason, we acquire images with an exposure time of 120 seconds, in order to let the signal accumulate and to reduce the number of readouts, thereby reducing the overall readout noise.

The dome is not yet adequately  dark, since there are some light leaks from the outside and some small light sources inside, such as LEDs or blinking lights from switches or electrical cabinets that are being progressively masked. For that reason, before and after every test execution we also acquire two dome background frames to estimate the background light reaching the focal plane that does not come from the CBP itself. Those frames are scaled to the test frames exposure time (the typical exposure time for LSSTCam images is 30 seconds) and subtracted from each one before combining them together. At the end, what we obtain in the processed images is solely the light coming out from the CBP and scattered/reflected from the M2 baffle.

All frames are then stacked together to form a full-band image that is useful to verify the effective presence of stray light; the full image processing workflow is described in Section \ref{SEC6}.

The in-dome tests were executed in August and December 2025.

\begin{figure}
    \centering
    \includegraphics[width=0.62\linewidth]{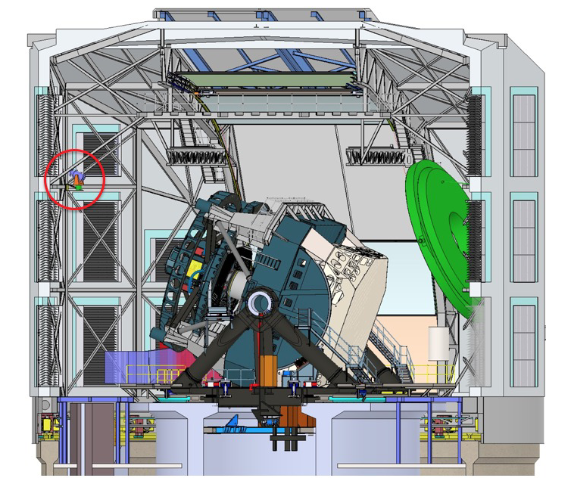}%
    \hfill
    \includegraphics[width=0.36\linewidth]{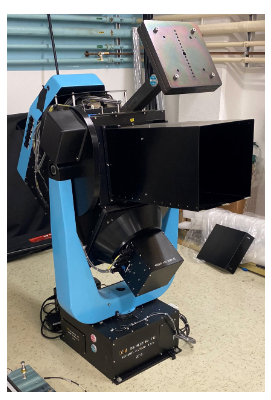}%

    \caption{Location of the Simonyi Survey Telescope and the CBP (left, circled in red) inside the Vera C. Rubin Observatory dome. {\textit{Up left:} The CBP during the last tests before being installed at its final location in the dome, as shown in right image. Credits: \cite{auxtel}}}
    \label{fig:cbpdome}
\end{figure}
\section{On-Sky tests}\label{SEC5}
The hints from ray tracing simulations, described in Section \ref{SEC3}, were confirmed from in-dome CBP tests, described in \ref{SEC4}. The next natural step was to observe the same effect on-sky to assess its impact on science images under realistic observing conditions, like those encountered during the 10-year LSST. The ray tracing simulations isolated the off-axis range in which the M2 baffle scattered light contribution is present as $10$°-$20$°, smaller off-axis angles lead to the build up of other contributions which are outside the scope of this paper. On-sky tests can verify if this effect is negligible or not and if it can compromise the compliancy to scientific requirements.

The test procedure foresees: placing a light source at the specific off-axis angle where the scattered light is maximized and seeing if the effect is detectable in the LSSTCam images. The first inspection is done visually, trying to spot gradients in the image where simulations and CBP tests suggest the light should fall in the focal plane. Other more accurate studies include some data reduction steps such as image stacking and/or stellar object removal. In such a wide field of view, it is not easy to spot a smooth gradient and distinguish it from other possible effects. The expected outcome from the on-sky tests is an increase of light in the outskirts of the image, where the vignetting effect is also strongly present, and the composition of the two can be misleading. Also, having a focal plane composed of two different detector populations makes it harder to identify smooth, faint gradients in the raw images. For all these reasons, a careful strategy of target selection is required; one must take into account both the spectral energy distribution (SED) of the celestial source used for the test and the M2 baffle reflectance and scattering properties at different wavelengths, and the interplay between the two.

Based on on-sky visibility, from the location of the Observatory \footnote{El Peñón peak of Cerro Pachón in Chile, 2647m above sea level, at coordinates 30°14'41"S, 70°44'58"W} , at the moment of test planning, the sources identified as useful for investigating the phenomena described in this paper were:
\begin{itemize}
    \item Antares star - HIP 80763 - Because of its high near-infrared emission visible in band $z$ and $y$, where the black anodized Aluminum reflects more light. $V_\mathrm{mag}=0.91$
    \item Alpha Centauri star - HIP 71683 - Because of its intrinsic high luminosity and frequent visibility from the Observatory site. $V_\mathrm{mag}=0.01$
    \item Spica star - HIP 65474 - because of its relatively high apparent magnitude, but mostly because it is far from the galactic plane and therefore in a sky region with lower nebulosity, thus faint surrounding effects are easier to detect. $V_\mathrm{mag}=0.97$
    \item The Moon thanks to its large apparent magnitude being brighter than any other object in the night sky. $V_\mathrm{mag}=-9.59$ at 30\% phase.
\end{itemize}

To maximize the possibility of observing visual affects from M2 baffle scattered light, all the designed tests were performed in $z$ band, where the combination of stellar light intensity and material reflectance is higher and at the off-axis. Since it is always mounted, using the $z$ filter gives more relaxed constrains to observability windows, unlike the $y$ filter which is alternated with the $u$ one.

The first test attempts were executed on-sky, targeting the three stars aforementioned. For a different set of reasons, the preliminary on-sky test failed to produce high-fidelity proof of the presence of scattered light coming from the M2 baffle; results from the three stars were not conclusive. This happened because of the complex overall background in the sky images; it is not possible to confirm whether the observed light comes from the M2 baffle or not. It was mainly a problem of contrast; the scattered light was too faint to be observed by eye and sky nebulosity was often predominant. To boost contrast, we moved our effort to use the Moon as a light source.

The idea of using the Moon as a source is not to measure the effect produced by the Moon itself because, during the nominal operation conditions for the survey, it is planned to have an avoidance radius around the Moon larger than the maximum off-axis distance at which the scattered light from the M2 baffle reaches the focal plane. So the choice of using the Moon is for demonstrative purposes only, to show with the minimum amount of image processing the presence of scattered light from the M2 baffle in accordance with the simulations. There are two kinds of tests executed with the Moon as a light source:
\begin{enumerate}
    \item Observations in $r$, $i$, $z$ bands at fixed off-axis distance to study the color dependence of the phenomena
    \item Raster observations in the most affected band at different off-axis angles between 14° and 20° \footnote{Going closer than 14° as we did with the CBP starts to be dangerous for the camera, if there is some unexpected light path that brings moonlight to the focal plane, this can saturate the detectors and generate risks for the related electronics.}
\end{enumerate}

The tests were executed in a wide time span: the first tests using the stars were executed in May 2025, while those using the Moon in April/May 2026 were executed after a serendipitous, unintentional close-up with the Moon between $14$° and $20$° in November 2025.

    \section{Retro-fitting and extrapolation} \label{SEC6}

Data obtained from on-sky tests and from in-dome tests belong to two distinct domains. What we need is a way to make them comparable in order to verify that they are mutually consistent. The two datasets have substantial differences both in the optical beam footprints and photometric properties. The exposure time of the image frame and the number of stacked images play an important role when trying to compare the results of different tests. Earlier, we said that the CBP is capable of emulating a star-like collimated beam for calibration purposes, but it is not capable of emulating its spectra and its footprint on the primary mirror. For these reasons, we need to do some extrapolations from one test configuration to the other. 

The acquired data can be divided into three groups:
\begin{itemize}
    \item In-dome CBP frames
    \item On-sky frames with stars as stray light source
    \item On-sky frames with the Moon as stray light source
\end{itemize}
In this section, we discuss the procedure followed to relate these three different datasets, in order to have a broad and complete understanding of how the system behaves.

\begin{wrapfigure}{r}{0.5\textwidth}
    \centering
    \includegraphics[width=0.48\textwidth]{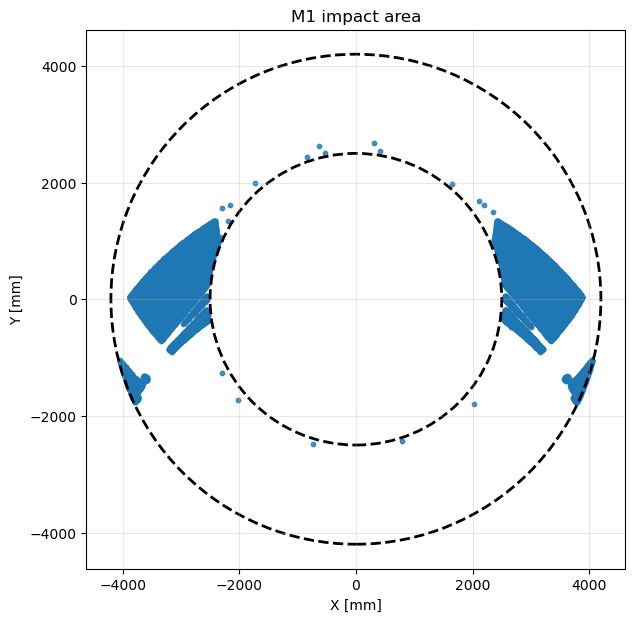}
    \caption{We reproduced the on-sky test scenario by shooting one million rays in a \texttt{Zemax} non-sequential ray tracing simulation, by saving the rays path and filtering them for the impact sequence M1-M2 baffle-M3-LSSTCam optics-focal plane, we obtained a map showing where rays hit the primary mirror. This image shows the rays footprint of a light beam 16 deg off axis.}
    \label{fig:impact}
\end{wrapfigure}

For the current study, the entrance pupil of the telescope can be approximated as the flat area of the annulus primary mirror, which has an outer radius of $4.2\mathrm{m} $ and an inner radius of $2.5\mathrm{m}$, resulting in an annulus of area $\approx 35\mathrm{m^2}$, the equivalent of an unobstructed $6.7\mathrm{m}$ diameter mirror. An on-axis source illuminates the whole pupil; therefore, its total intensity in the acquired image, after a fixed exposure time, depends on the telescope's collective area. The same relation does not hold for strongly off-axis sources, such as the ones we are investigating in this study, because the concept of collective area of the telescope is defined only for sources inside the nominal field of view. For off-axis sources, all the obstructions in the telescope assembly prevent part of the incoming beam from reaching the primary mirror, and this results in a footprint on the primary mirror smaller with respect to on-axis ones. Including everything on top of the telescope assembly in a ray tracing software makes it possible to find the impacted area by the light coming from off-axis sources. Figure \ref{fig:impact} shows the impact area on the primary mirror for a source located 16° off-axis. By means of automatic image analysis scripts, it is possible to compute the fraction of primary mirror area where the light is effectively collected when in the presence of highly off-axis sources.

\begin{table}[h]
    \centering
    \begin{tabular}{|c|c|c|c|}
        \hline
           &                    12°        & 16°        & 20° \\
        Total area                &  $35.7\mathrm{m^2}$ & $35.7\mathrm{m^2}$    & $35.7\mathrm{m^2}$ \\
        Stellar illuminated area  & $3.6\mathrm{m^2}$  &  $2.4\mathrm{m^2}$   &   $0.83\mathrm{m^2}$\\
        CBP illuminated area      & $0.045\mathrm{m^2}$ & $0.045\mathrm{m^2}$ & $0.045\mathrm{m^2}$ \\
        fraction CBP/Star                  &  $1.25\%$ &  $1.87\%$ &  $5.4\%$ \\
        \hline
    \end{tabular}
        \vspace{2mm}

    \caption{The evolution of the light footprint as a function of the off-axis angle for a stellar object and the CBP beam. The star footprint decreases as the off-axis angle increase, thus causing the ratio with the fixed CBP beam width to increase.}
    
    \label{tab:area}
\end{table}
With the same approach, we can also compute the impact area on the primary mirror from the CBP. We know that by design,  the CBP can produce a $\sim24cm$ diameter collimated beam \cite{cbp}. We rebuilt the CBP inside our ray tracing software and verified the aforementioned dimension; from the analysis, it results in $23\mathrm{cm}$ wide beam.

When comparing the CBP beam bandwidth with the SED of a real star or the Moon, the extrapolation is simpler than for the light footprints comparison. Since the CBP is highly monochromatic, we can assume it has a bandwidth of $\sim 1\mathrm{nm}$ when propagating light. As previously discussed, the main differences between this testing configuration and a real observing scenario are both the polychromaticity of a real celestial source and also the fact that in the CBP tests, we are not sampling each possible wavelength inside each filter window because of testing time limitations.

With the CBP we sampled bands $i$, $z$, $y$ where the reflectivity of the anodized Aluminum is higher, at wavelength steps of $20\mathrm{nm}$ inside each band with a  $\sim1\mathrm{nm}$ bandwidth. Taking the full window width for each filter and dividing it by the number of sampled wavelengths with the CBP, one can find another corrective factor to be multiplied by the CBP counts in order to compare them with the on-sky counts produced by the moon or one of the stars used for the tests (Table \ref{tab:star_ct}). This statement is based on the assumption of a gray profile for the baffle reflectance inside each filter. We only have one measurement inside  $i$, $z$ and $y$  band each, therefore we assume this value for the full band. The results shown in Figure \ref{fig:bafref} motivate this assumption, since inside $i$, $z$ and $y$  bands the reflectance is almost constant.

Another difference between the two datasets is the image frame exposure time. On-sky images are always acquired with a 30 sec exposure time since this is the standard for LSSTCam operations, and we kept this value fixed to resemble real operations. During the CBP test campaign instead, to maximize the signal-to-noise ratio (SNR), we increased the exposure time to 120 seconds, acquiring 4 images for each sampled wavelength, thus resulting in 480 seconds of exposure for every single sampled wavelength. This effect has to be taken into account to normalize the counts to a 30-second exposure. This specific factor can change depending on how one decides to process the single frames, summing the 120 second long frames results in having a total of 480 second exposure times, while mediating them, instead, result into an equivalent frame of 120 seconds. We decided to sum the images in each quadruplet and then subtract from them a master dark, scaled by a factor of 4 to produce the single wavelength image containing only the scattered light component. With the CBP, the scattered signal is very faint, so the dark/background subtraction is a necessary step to isolate the gradient produced by the scattered light we are investigating, which would be lost in the noise otherwise. Also, to isolate the energy coming from the scattered light, it is necessary to remove environmental contributions as much as possible.

Other factors to be taken into account when trying to compare on-sky and in-dome tests are the absolute throughput of the system for celestial sources and the different flux density between CBP and astronomical source. Since the CBP is located inside the dome, it is not affected by atmospheric absorption but only by the telescope's
intrinsic optical throughput.  This means that a factor accounting for the missing atmospheric absorption must be added to our estimations. Regarding the flux density, we decided to add in our relation a theoretical ratio between CBP flux density and a star to which we want to compare. This multiplicative factor takes into account the higher flux coming from the CBP with respect to on-sky sources.

Taking account of the factors discussed above, an approximated way to translate the CBP results into on-sky predictions is to scale the CBP images following the relation:
\begin{equation}
   \mathrm{IMG}_\mathrm{star}= \mathrm{IMG}_\mathrm{CBP} \times \mathrm{A} \times \mathrm{W} \times \mathrm{T}_\mathrm{exp}  \times \mathrm{TP} \times \mathrm{F}_\mathrm{SED}
    \label{eqn:cbp}
\end{equation}
 where the factors represent, respectively:  $\mathrm{A}$ the correcting factor for the collective area as shown in Table \ref{tab:area};   $\mathrm{W}$ the factor relative to the wavelength sampling which is 16.6 for  $i$ and $y$  bands and 20 for the $z$ band if we start from multi-wavelength stacked images;  $\mathrm{T}_\mathrm{EXP}$ the different exposure times of the stacked image result from all the acquired frames; $\mathrm{TP}$ the measured overall throughput of the telescope combined with the atmosphere;  $\mathrm{F}_\mathrm{SED}$ the ratio between the star flux density and the CBP flux density. 

 The relation in Equation \ref{eqn:cbp} relies on a lot of assumptions and has to be used as last resort when trying to compare CBP in-dome acquired data with hypothetical on-sky observations made using a specific star  as light source. Instead of using the CBP images as quantitative tracers for the expected counts on-sky, we decided to keep the CBP data as a reference point to compare with the simulations. This allowed us to verify that the predicted light path in Figure \ref{fig:model} exists and produce an increased background level in the same region of the focal plane that simulations suggest. Since we have on-sky observations from both stars and the Moon, which makes the feature easily detectable, it is simpler to compute the expected counts going from the Moon observations to stars, rather than from the CBP frames to stars. Making the Moon-to-stars comparison relieves us from using the factors referring to impact area, wavelength sampling, throughput and exposure time. We simply have to rescale what we see in Moon images through the different flux density between the Moon and the stars. Only the factor $\mathrm{F}_\mathrm{SED}$ remains in this scenario. Under the assumption of proportionality between flux and instrumental counts collected by the LSSTCam detectors \cite{lsstcam} \cite{lsstcam2} \cite{lsstcam3} , confirmed by the non-saturation of the detectors during our tests, we can write down the following equation:
\begin{equation}
    \mathrm{Flux}_\mathrm{moon}:\mathrm{Flux}_\mathrm{star}=\mathrm{Count}_\mathrm{Moon}:\mathrm{Count}_\mathrm{star}
\end{equation}
Where the same also holds between the Moon and CBP. Knowing the flux of the Moon at a specific phase and the flux of a specific star, after isolating the increased background contribution observed in on-sky test frames seen in Figure \ref{fig:moontest}, one can compute the amount of counts related to scattered light a star should produce simply by:
\begin{equation}
    \mathrm{Count}_\mathrm{star}=\frac{\mathrm{Flux}_\mathrm{star}}{ \mathrm{Flux}_\mathrm{moon}}\cdot   \mathrm{Count}_\mathrm{Moon}
    \label{eqn:ctstar}
\end{equation}

\noindent The same relation but used with the known CBP counts can also be helpful to estimate the effective CBP flux in terms of $\mathrm{erg/s/cm^2/}$\AA.

\begin{figure}
    \centering
    \includegraphics[width=0.45\linewidth]{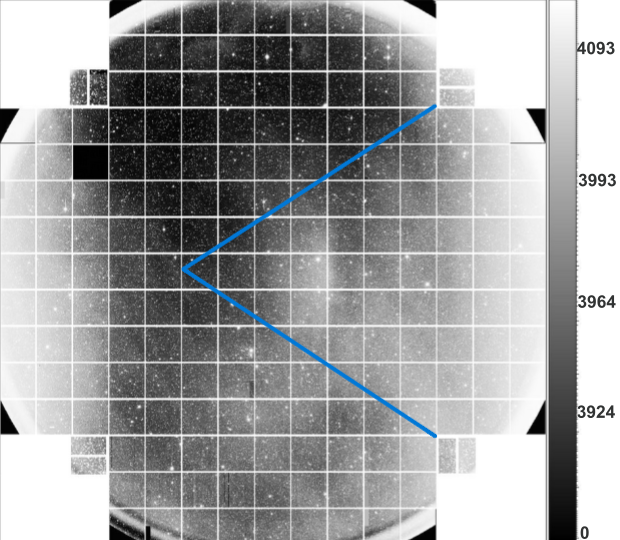}
    \includegraphics[width=0.45\linewidth]{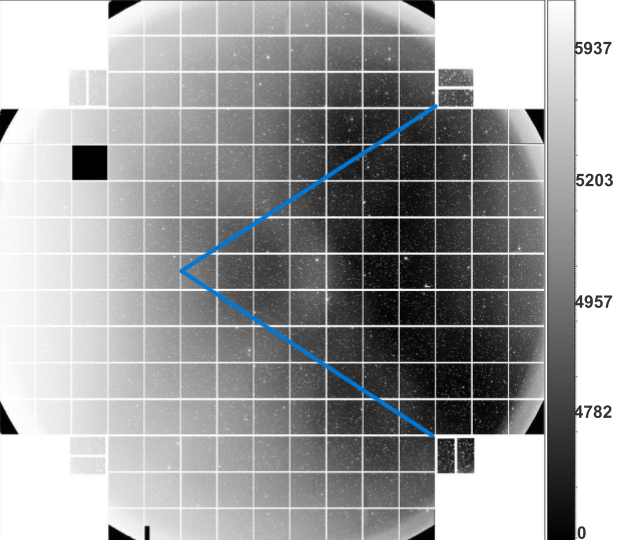}
    
    \includegraphics[width=0.45\linewidth]{images/moontestZ.png}

    \caption{From top left to right, the upper row shows  $r$ and $i$, while the bottom row shows $z$ bands on-sky image acquired placing the Moon 16° off-axis. A symmetric triangle-shaped pattern can be observed in the left part of the image, resembling what the simulations suggested for the feature shape on the focal plane. The level of counts in the affected region respect the expected color dependence already discussed. On the images a triangle shaped geometry is manually added to highlight the region affected by the scattered light, the \textit{Pacman-shaped} geometry outside the triangle is where we expect to see the light in the focal plane.}
    \label{fig:moonraster}
\end{figure}

Regarding Equation \ref{eqn:ctstar}, we made our estimations in $z$ band with a Moon phase of 30\%, which is the same as the observing setup. Starting from a full Moon magnitude of $-12.55\mathrm{mag}$ \cite{moon}, we estimated the strongest M2 baffle scattered light contribution to the background is $\sim 6000 \mathrm{e^-}$, resulting from the difference between the median counts in an affected region and in an unaffected one in the same image. In particular we encountered the highest contamination in the frame at 14.5° off-axis acquired during the rastering test, top left image in Figure \ref{fig:moonraster}. Using the $z$ band fluxes of the aforementioned stars in Section \ref{SEC5}, we obtained the estimations of the diffuse background coming from the light scattered by the M2 baffle shown in Table \ref{tab:star_ct}.

\begin{table}[h]
    \centering
    \begin{tabular}{|c|c|c|c|c|c|}
    \hline
        Star           & V mag & Spectral type & $z$ mag & $\mathrm{e^-}_\mathrm{z}$ \\
        Alpha Centauri & 0.01  & G2V           &  -0.8  &  0.7  \\
        Antares        & 0.91  & M1.5I         &  -2    &  6.5  \\
        Betelgeuse     & 0.42  & M2I           &  -2.8  &  4.8  \\    
        Spica          & 0.97  & B1V           &   0.2  &  0.3 \\
   
    \hline

    \end{tabular}
    \vspace{3mm}
    \label{tab:star_ct}

    \caption{The recomputed magnitudes and flux density in LSSTCam $z$ band for the stars used in the on-sky test campaign. Together with the Moon flux density in $z$ band and the actual on-sky counts in the acquired frames these values are used to compute the equivalent ADU the stars should produce on the focal plane due to the M2 baffle scattered light contribution in a 30 second exposure. Single stellar component is negligible.}
\end{table}
 
    \section{Results}
\subsection{Simulations results}\label{sec71}
The M2 baffle is made of a type of black anodized Aluminum with a reflectivity close to 30\% in band $z$ and $y$. As a first approximation, this reflectivity has been applied in the simulation models by treating the M2 baffle as a mirror with that 30\% reflectance. Preliminary simulations showed a double-triangular shaped pattern, shaped by the light vanes of the M2 baffle itself, that can be seen in Figure \ref{fig:fotobaf}. 
\begin{figure}
    \centering
    \includegraphics[width=0.49\linewidth]{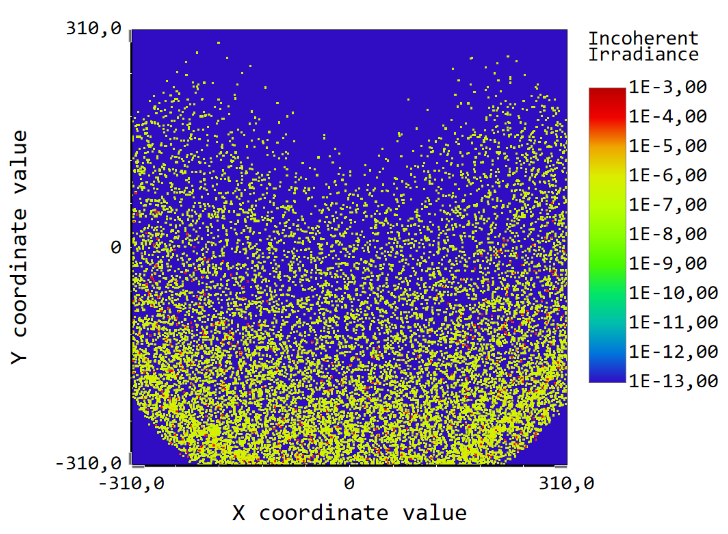}
    \includegraphics[width=0.49\linewidth]{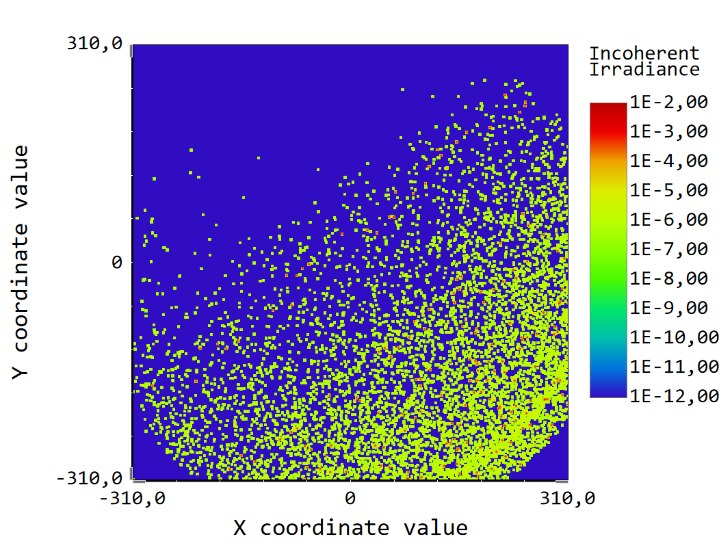}
        \caption{\textbf{Left:} Full focal plane simulation for a $16$° off-axis beam illuminating the entire telescope. The corresponding beam footprint is highlighted in Figure \ref{fig:impact}. \textbf{Right:} For clarity, pattern produced by light from the right lobe of the footprint in Figure \ref{fig:impact}. The two components overlap at the focal plane center, forming the quasi-triangular shape also seen in the on-sky frames.}
    \label{fig:focalplane}
\end{figure}

\begin{wrapfigure}[24]{r}{0.5\textwidth}
    \centering
    \vspace{-4pt}
    \includegraphics[width=0.48\textwidth]{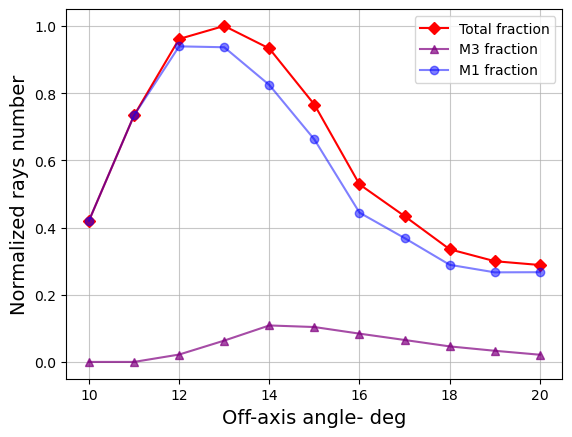}
        \caption{The result  of the simulated rastering between 10° and 20° is the relation showed here where we can see how the number of rays impinging on the primary mirror change with respect to the off-axis angle considered. In simulations we kept the amount of shot rays fixed at 50 million, such that any change in the number of incoming rays on M1 that bounce on M2 is purely geometrical. The number of rays refers to the ones reaching the LSSTCam detector. The red line (Total fraction) is the sum of the other two contribution representing the two identified stray light paths. }
    \label{fig:ctsangle}
\end{wrapfigure}

The first results are used as a hint to design in-dome and on-sky tests, which are then compared to simulations. The latter are refined to match the real case as best as possible. In this way, we are sure about the truthfulness of our model.

By rastering a stellar source at different off-axis angles, the simulations suggest that the peak of the scattered light background is around $13$°, while the whole inspected range spans from $10$° to $20$°, as shown in Figure \ref{fig:ctsangle}. The simulations let us investigate the light path, its geometry, and the scattered background light shapes on the detector plane. To recover the energetics of the phenomena, studies about the effective intensities and real observations are required; the results from our on-sky and in-dome campaigns are described in the following subsections. 

The simulations indicate the off-axis range where the scattered light should manifest, the light path to materialize it, and the expected light distribution on the LSSTCam focal plane.

\subsection{In-dome campaign results}
The in-dome test campaign, discussed in Section \ref{SEC4}, revealed that the stray light path identified in the simulations is indeed present and it materializes during the test as a diffuse halo in the corner of the images. Using the CBP, we explored the wavelength bands where the reflectance of the black anodized Aluminum composing the M2 baffle is higher; bands $i$, $z$ and $y$  for a window 300nm wide, from 720nm to 1020nm.

\begin{figure}
    \centering
    \includegraphics[width=\linewidth]{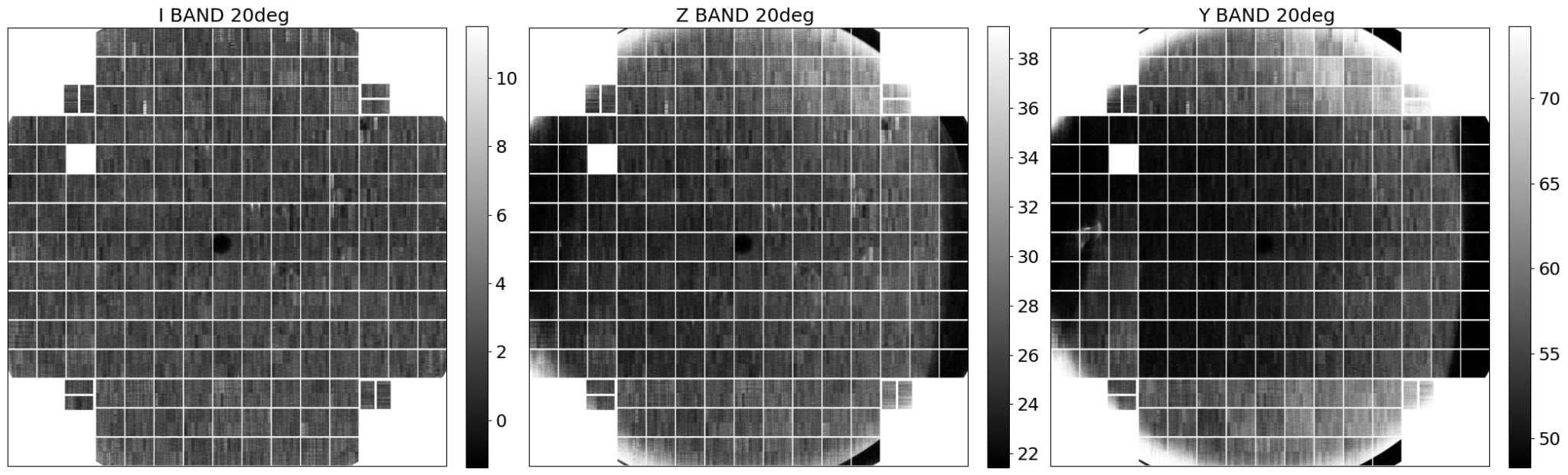}
    \includegraphics[width=\linewidth]{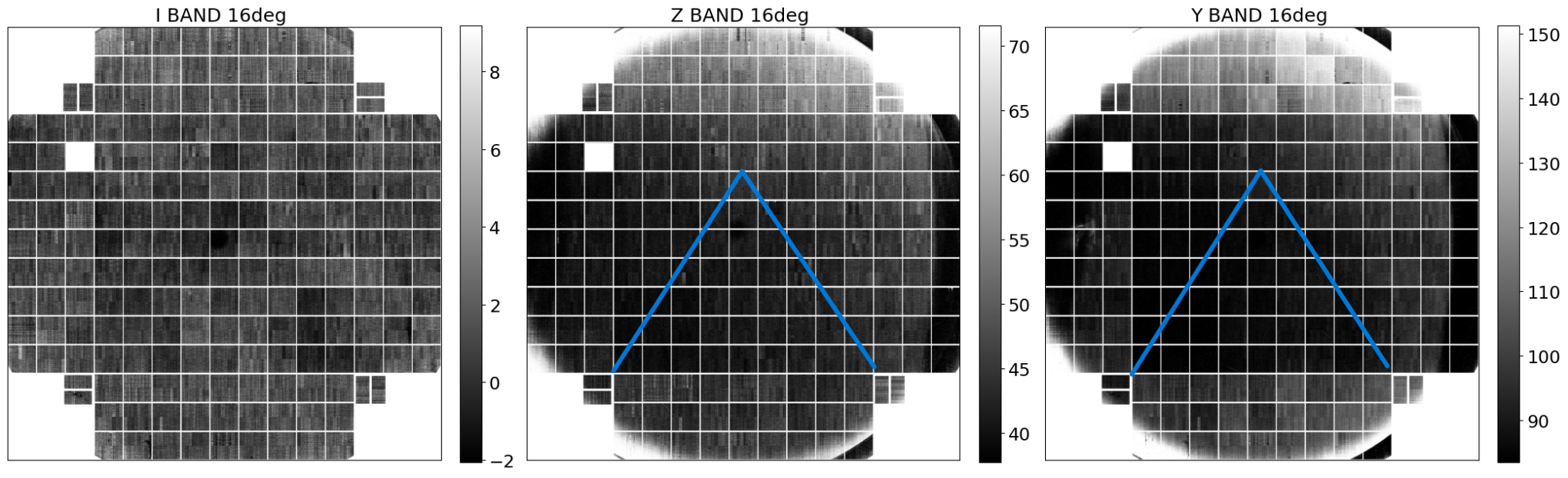}
    \includegraphics[width=\linewidth]{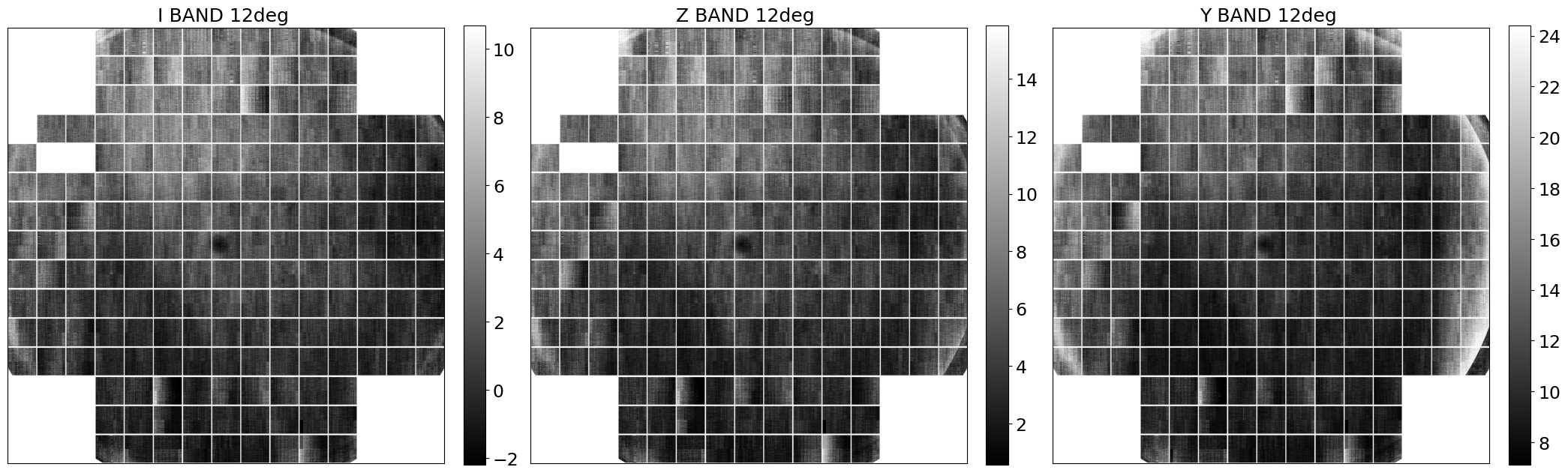}
    
    \caption{This mosaic contains the data acquired during the in-dome campaign. The raw data have been processed as discussed in  \ref{SEC6}. The two black arrows point to the region with  higher counts at $16$° off-axis which are due to the scattered light from M2 as suggested by the simulations. The observed region with an increased count level closely matches the simulation of the one-lobed source shown in the right panel of Figure \ref{fig:focalplane}, both in terms of position and overall geometry. The $y$ band which is the most affected by the scattered light is also the one that suffers more from the discontinuity between the two detector families in the focal plane.  This makes it easier to spot the diffuse emission in the $z$ band (central panel) instead. In the two most affected images, where it is easier to sobserve the scattered light, a triangle shaped geometry is manually added to highlight the region affected by the scattered light and to easily link the focal plane feature geometry coming from the CBP and from the on-sky tests. The \textit{Pacman-shaped} geometry outside the triangle is where we expect to see the light in the focal plane, but since the CBP is only able to illuminate a small portion of M1 in these images we only observe half of the fenomena as in the right panel of Figure \ref{fig:focalplane}}
    \label{fig:cbpresult}
\end{figure}

Thanks to the image processing discussed in Section \ref{SEC4}, the scattered light coming from the M2 baffle is also visible in the three panels in Figure \ref{fig:cbpresult}, where we show the stacked and background-subtracted images at the three sampled off-axis angles. The higher counts region in the upper/right upper region of the image, visible thanks to the scaling and the color map choice, represents the increase of light coming from the M2 baffle scattered light. It is the in-dome counterpart of what is shown in the right panel of Figure \ref{fig:focalplane}; the CBP is able to illuminate only one of the two footprint's lobes, therefore the results consist of a one-sided feature corresponding to the illuminated lobe. Between simulations and CBP images, there is a rotation factor because of the different pointing method used with the CBP with respect to the model. In \texttt{Zemax}, we are allowed to move both the source and/or telescope independently, while in reality, the CBP is fixed and can only move azimuthally, hence a specific pointing is required to achieve the desired off-axis angle.

At first glance, the results shown in Figure \ref{fig:cbpresult} seem to be in contrast with the trend of the number of rays observed in the rastering simulations in \texttt{Zemax}, Figure \ref{fig:ctsangle}. The explanation is that they actually sample and show different behaviors. As already discussed, the CBP is able to illuminate a small portion of the entrance pupil of the telescope, while in the simulations we used a bigger source to resemble the illumination from a celestial body. Possible explanation for that observed contrast can be that the asperical shape of M1 and the small, but present, divercence of the CBP beam end up into having a slightly different optical path that hit differently the M2 baffle, therefore produce less counts in the focal plane. This tension betwee simulations and CBP tests was also a strong motivation to go for on-sky tests to effectively observe how the system respond to real off-axis sources.

Regarding the image frames in band $i$, we see a very low level of counts and negligible background contamination.  This is due to a combination of different factors: the low emissivity of the CBP laser and the lower reflectivity of the M2 baffle itself. The absence of scattered light in the in-dome test in I band is expected from previous reflectance measurements on the M2 baffle and is discussed in Section \ref{SEC2}. From the colorbar levels in the $z$ and $y$  band plots, we can see an increase in the total counts related to the scattered light. This result is expected, and it can be explained with the same reflectance curve and by the fact that the CBP laser is more intense in those wavelengths. Thanks to the indications of the simulation, we succeeded in reproducing the stray light path from the real M2 baffle into the LSSTCam. These results allowed for the understanding and the confirmation of the scattering process, and they motivated the need for on-sky tests with both stellar sources and the Moon. 

\begin{figure}
    \centering
    \includegraphics[width=0.45\linewidth]{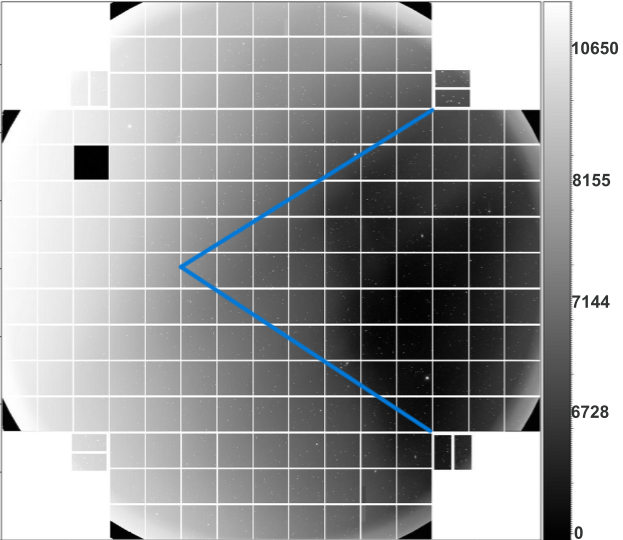}
    \includegraphics[width=0.45\linewidth]{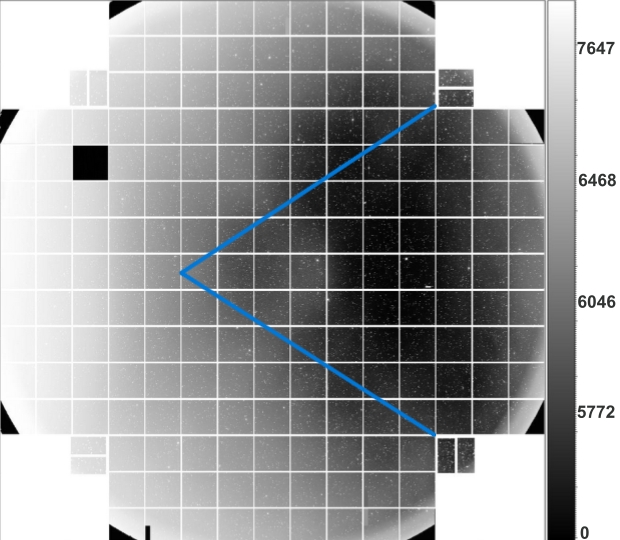}
    \includegraphics[width=0.45\linewidth]{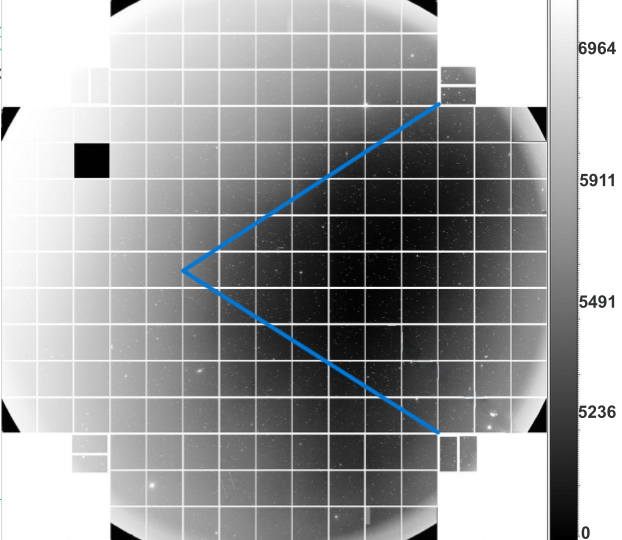}
    \includegraphics[width=0.45\linewidth]{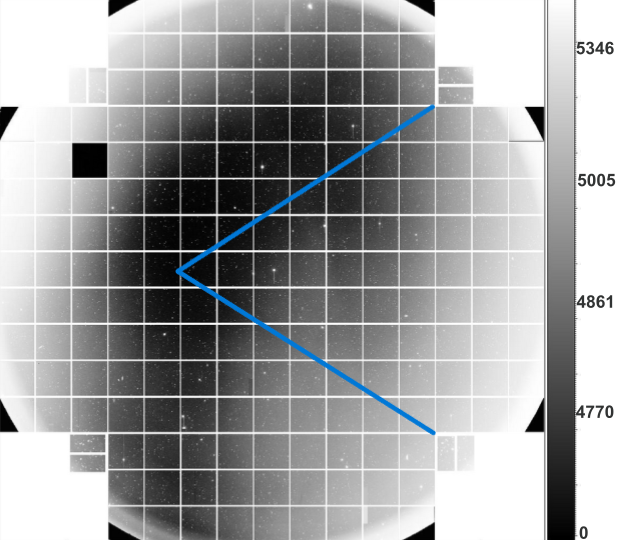}

    \caption{The top panel from left to right shows, respectively, the images acquired at $14.5$° and $16$° off axis in $z$ band.  The bottom panel shows images acquired at $17$° and $19$° in $z$ band during the same test. The evolution of the color bar limits helps understanding how the intensity of the phenomena changes with the off-axis angle. As suggested from simulation and discussed in Figure \ref{fig:ctsangle}  the peak intensity is around $13$°. On the images a triangle shaped geometry is manually added to highlight the region affected by the scattered light, the \textit{packman-shaped} geomatry outside the triangle is where we expect to see the light in the focal plane.}
    \label{fig:moontest}
\end{figure}

Scaling the CBP effect to a stellar case requires some assumptions and extrapolation of the available data. We show that the number of counts in the focal plane due to the increased background from the M2 baffle scattered light, for an astronomical source as bright as the CBP, is around 1000. This means in Equation  \ref{eqn:cbp}, $\mathrm{F}_\mathrm{SED}=1$. Comparing the extrapolated result with the on-sky test is useful to verify that the set of assumptions made to derive it was correct. In the next subsection, Figure \ref{fig:moontest} shows that the left side of the image, which is contaminated by the scattered light coming from the Moon, has almost 1000 more counts than the right part of the frame, which is clean.

The in-dome test campaign resulted in a practical confirmation of the reflectivity of the M2 baffle, and it showed that this contribution is appreciable if the off-axis light source is bright enough to overcome the general background level in the frames. It also confirmed the fact that the effect is color-dependent, becoming more relevant in the redder wavelength. There is still a residual degeneration in the results between the intrinsic laser emission at different wavelengths and the M2 baffle reflectivity. On-sky tests are also helpful to break this degeneracy because the sources are well known, and we can select a celestial body to fit our requests in terms of spectra.

\subsection{On-sky campaign(s) results}
In the previous sections, the testing procedures were explained, as well as the analysis used to motivate the on-sky campaign using the Moon as a light source. As already mentioned, the on-sky campaign using stars as a light source failed to confirm the presence of the increased background due to the M2 baffle scattered light.  This is due to the intrinsic faintness of the effect and the complexity of the sky background itself, in particular at the edges of the field of view, where we expect to see the most structured contribution. The increase in ADU produced from stars is discussed in Table \ref{tab:star_ct}.

Frames acquired using the Moon as an off-axis light source succeeded in confirming the results obtained from both simulations and the in-dome campaign. Despite some rotational offsets, we can see a very clear agreement between simulations, CBP stacked frames and on-sky images. The on-sky tests focus on replicating the color dependence of the increased background and scanning the range of angles where the contamination happens. The light path and the geometry causing the light to hit the M2 baffle and then scatter onto the LSSTCam focal plane are better characterized by simulations and CBP tests.

The off-axis scan to sample the whole range where the stray light is present comprehends 4 different data points, from $14.5$° to $19$° in steps of $1.5$°. The results showed a variation in the number of median counts in the affected regions comparable to the expected counts from the simulations, Figure \ref{fig:ctsangle}.

    \section{Conclusions}

We developed three different and independent methodologies to evaluate the impact of scattered light from the M2 baffle: ray tracing simulations with \texttt{Zemax}, CBP testing, and on-sky testing using bright stars and the Moon. The investigations led to the retrieval of consistent evidence of the M2 baffle scattered light behavior and impact on the LSSTCam images.
It is important to note that the original optical design assumptions for the M2 baffle were based on ray-tracing studies that considered only light propagating through the nominal clear aperture under fully operational observing conditions including a fully deployed LWS. In these early analyses, large off-axis incidence angles were not extensively explored and were more focused on determining the geometry and location of the different baffles rather than their own net contamination. Since a clear advantage of a coated surface was not obvious during the design and fabrication of the M2 baffle, and painting the large number of vanes would be difficult, optical surface treatments were not incorporated. This design context motivated the decision not to apply a high‑performance, low‑reflectance coating (e.g., Aeroglaze Z306) to the M2 baffle.
In the previous sections, it has been demonstrated and discussed how the M2 baffle can produce an increased light background due to the light being scattered on its surface when illuminated by a $12$ to $20$ degrees off-axis source. This effect appears as a diffuse halo extending over roughly $2/3$ of the focal plane; it is formed by two half-disks rotated by 90° around the optical axis; their overlap creates a triangular region where the effect is more pronounced. The final shape is referred to as being "Pacman-shaped".

Analysis and measurements of the optical properties of the M2 baffle suggest that the expected scattered light artifacts in the images are color-dependent. This is confirmed by both the CBP test results in Figure \ref{fig:cbpresult} and the on-sky images shown in Figure \ref{fig:moontest}, where the contamination is higher in the redder bands as suggested by the higher reflectivity measurements retrieved from the M2 baffle and shown in Figure \ref{fig:bafref}. 

We used the Moon as a light source to boost the contrast between the scattered light from the M2 baffle and the nominal sky background, even though during operations the current minimum angular distance is $30$° \cite{Survey_Constraints} as best practices to avoid scattered light. If the limits hold, then there shouldn't be any contamination coming from the Moon regarding the specific artifact object of this paper. The pure contribution of the scattered light from the M2 baffle decays beyond an angular separation of $20$°. The larger off-axis component showed in Fig \ref{fig:ctsangle} will be blocked by the Light Wind Screen (LWS) when it will be fully operative. This screen will block all the light coming from the surrounding of the imaged sky area in a square region of $\pm 10$° in each direction around the pointing location. The maximum off-axis angle allowed to enter the shutter will be $\sim 14$°, thus the peak of intensity could still leak into the telescope, but the tail from $14$° on will be blocked. This prospective shifts the discussion about the -future- most impactful contributor to stars and planets rather than the Moon itself.  What remains as a possible source of contamination are bright stars, the red ones in particular. From the Rubin Observatory location, only 45 stars with magnitude brighter than $2 \mathrm{V}_\mathrm{mag}$ are visible, among those only 10 \footnote{Resulting query from Vizier/SIMBAD HIP 21421, HIP 24608, HIP 27989, HIP 37826, HIP 41037, HIP 61084, HIP 69673, HIP 71681, HIP 71683, HIP 80763 } can be considered red, i.e. they have a B-V color $>0.7$ resulting in a $z$ magnitude $<0.5$. In Table \ref{tab:star_ct} we show the expected counts coming from some of the brightest stars in the sky, both red and not. During the on-sky campaign, even the brightest stars were not able to produce an appreciable number of counts; the majority of them are expected to produce a level of counts comparable or smaller to the readout noise of the CCDs in the focal plane \cite{readnoise} \cite{lsstcam}. Only the brightest ones can reach $4\mathrm{e^-} - 6\mathrm{e^-}$ per pixel per $30\mathrm{s}$ frame; this scenario can be more problematic in the case of structured jumps and edges. Data processing pipelines can deal with smooth gradients but can end up over or under subtract signal close to sharp and abrupt variations of counts. A clear example of this sharp-edge tentative removal is observed in other stray light features \cite{wagner2026} \cite{gabri2026}. We did try to go further in the image processing using the nominal pipelines, even when the moon was the culprit light source, the increased gradient in some portion of the images, due to the M2 baffle scattered light, was removed without any evident signals of over or under subtraction. This led us to conclude that if the Moon itself, which produces hundreds of $\mathrm{e^-}$ is easily removed, then the very limited contamination coming from the stars is nothing to worry about, as long as it is not producing sharp edges.   

In summary, we empirically verified the presence of two paths of stray light coming from light scattered off the M2 light baffle using red stars and the Moon as light sources at different off-axis angles. We extrapolated the equivalent intensity of the scattered light in the LSSTCam focal plane coming from the Moon, we translated it to the contribution of any star. We found that the stellar contamination is minimal and comparable to the readout noise of the LSSTCam. Depending on the minimum angular distance from the Moon that will be chosen for the LSST scheduler, the M2 baffle re-coating with a low-reflectance, high-scattering paint might be required. If the survey requires a minimum angular Moon distance $<$ 20°, a re-coating of the M2 baffle is highly recommended.

    \section*{Acknowledgments}
The authors acknowledge the contribution of the USC A of the INAF Scientific Directorate to the program "Italian participation in the Rubin LSST project". The authors acnowledge Matteo Munari for helping obtain the scattering models on Aeroglaze $\mathrm{Z}306$.
This material is based upon work supported in part by the National Science Foundation through Cooperative Agreements AST-1258333 and AST-2241526 and Cooperative Support Agreements AST-1202910 and 2211468 managed by the Association of Universities for Research in Astronomy (AURA), and the Department of Energy under Contract No. DE-AC02-76SF00515 with the SLAC National Accelerator Laboratory managed by Stanford University. Additional Rubin Observatory funding comes from private donations, grants to universities, and in-kind support from LSST-DA Institutional Members.

\bibliography{report} 
\bibliographystyle{spiebib} 

\end{document}